\documentclass[nonacm,manuscript,screen,table,sigconf,prologue,dvipsnames]{acmart}

\usepackage{algorithm}
\usepackage{algpseudocode}
\usepackage{cancel}
\usepackage{graphicx}
\usepackage{textcomp}
\usepackage{xcolor}
\usepackage{multirow}
\usepackage{tikz}
\usepackage{subcaption}
\usepackage{pifont}
\usepackage{listings}
\usepackage{fontawesome5}

\AtBeginDocument{%
  \providecommand\BibTeX{{%
    \normalfont B\kern-0.5em{\scshape i\kern-0.25em b}\kern-0.8em\TeX}}}
    
\newcommand{\xmark}{\ding{55}}%
\newcommand{\cmark}{\ding{51}}%
\newcommand{\hdagger}{\makebox[0pt][l]{$^\dagger$}}%
\DeclareRobustCommand{\circled}[1]{\raisebox{.5pt}{\textcircled{\raisebox{-.9pt} {#1}}}}%

\DeclareRobustCommand{\semeq}{\ensuremath\mathrel{\widehat{=}}}%
\DeclareRobustCommand{\nsemeq}{\ensuremath\mathrel{\cancel{\widehat{=}}}}%

\newcommand*{\TakeFourierOrnament}[1]{{%
\fontencoding{U}\fontfamily{futs}\selectfont\char#1}}
\newcommand*{\danger}{\TakeFourierOrnament{49}}

\setcopyright{none}
\copyrightyear{2023}
\acmYear{2023}
\acmDOI{XXXXXXX.XXXXXXX}

\acmConference[SC23]{The International Conference for High Performance Computing, Networking, Storage, and Analysis}{November 12--17, 2023}{Denver, CO}
\begin{document}

\title[FuzzyFlow]{FuzzyFlow: Leveraging Dataflow To Find and Squash Program Optimization Bugs}

\author{Phlipp Schaad}
\email{philipp.schaad@inf.ethz.ch}
\orcid{0000-0002-8429-7803}
\affiliation{%
  \institution{ETH Zurich}
  \city{Zurich}
  \country{Switzerland}
}

\author{Timo Schneider}
\email{timos@inf.ethz.ch}
\affiliation{%
  \institution{ETH Zurich}
  \city{Zurich}
  \country{Switzerland}
}

\author{Tal Ben-Nun}
\email{talbn@inf.ethz.ch}
\affiliation{%
  \institution{ETH Zurich}
  \city{Zurich}
  \country{Switzerland}
}

\author{Alexandru Calotoiu}
\email{acalatoiu@inf.ethz.ch}
\affiliation{%
  \institution{ETH Zurich}
  \city{Zurich}
  \country{Switzerland}
}

\author{Alexandros Nikolaos Ziogas}
\email{alziogas@inf.ethz.ch}
\affiliation{%
  \institution{ETH Zurich}
  \city{Zurich}
  \country{Switzerland}
}

\author{Torsten Hoefler}
\email{htor@inf.ethz.ch}
\affiliation{%
  \institution{ETH Zurich}
  \city{Zurich}
  \country{Switzerland}
}

\renewcommand{\shortauthors}{Schaad et al.}

\begin{abstract}
The current hardware landscape and application scale is driving performance engineers towards writing bespoke optimizations.
Verifying such optimizations, and generating minimal failing cases, is important for robustness in the face of changing program conditions, such as inputs and sizes.
However, isolation of minimal test-cases from existing applications and generating new configurations are often difficult due to side effects on the system state, mostly related to dataflow.
This paper introduces FuzzyFlow: a fault localization and test case extraction framework designed to test program optimizations.
We leverage dataflow program representations to capture a fully reproducible system state and area-of-effect for optimizations to enable fast checking for semantic equivalence.
To reduce testing time, we design an algorithm for minimizing test inputs, trading off memory for recomputation.
We demonstrate FuzzyFlow on example use cases in real-world applications where the approach provides up to 528 times faster optimization testing and debugging compared to traditional approaches.
\end{abstract}

\settopmatter{printacmref=false}
\renewcommand\footnotetextcopyrightpermission[1]{}

\maketitle

\section{Introduction}
Automatic compiler optimizations play an important role in getting good performance out of modern scientific applications.
However, the limited knowledge a compiler has over certain program parameters can cause overly conservative optimizations.
Applications are often further fine-tuned manually by performance engineers for specific input data or hardware architectures to exploit more potential performance gains.
Manual application of such repetitive optimization patterns can be time consuming, which is why performance engineers may provide compilers with purpose-built, more aggressive optimization passes to be applied to a program at a large scale~\cite{Briggs2022ChoosingAccuracy, Lattner2021, Kjolstad2017Taco:Kernels}.

While optimizing compilers are well-studied and tested tools, with a lot of research going into making sure their code transformations preserve program semantics~\cite{Leroy2006FormalAssistant, Leroy2009FormalCompiler, Kumar2014CakeML:ML}, the same does not hold true for custom optimization passes.
To apply such a custom transformation indiscriminately, it consequently has to go through rigorous testing.
Unfortunately, writing hundreds of test cases for a transformation is both labor intensive and may still miss crucial corner cases.

To avoid this issue, transformation rules themselves may be formally verified to ensure they preserve program semantics~\cite{Rinard2003CredibleCompilation, Leroy2009FormalCompiler, Necula1998TheCompiler, Pnueli1998}.
However, the correctness of certain transformations may depend on specific parameters or program inputs~\cite{Briggs2022ChoosingAccuracy}, making it difficult for this technique to detect issues related to changing program conditions.
Checking for semantic preservation through symbolic execution and concolic testing~\cite{Hathhorn2015DefiningC, Cadar2008}, or fuzzing and random testing~\cite{Boujarwah1997CompilerAssessment, Yang2011FindingCompilers, Manes2018} addresses this issue and allows transformations to be checked under various conditions instead.
Unfortunately, this strategy is often impractical in the context of long-running scientific computing applications such as numerical weather prediction or machine learning, where a single program execution can consume thousands of node hours and problem sizes may be fixed.

\begin{figure}[t]
  \includegraphics[width=\linewidth]{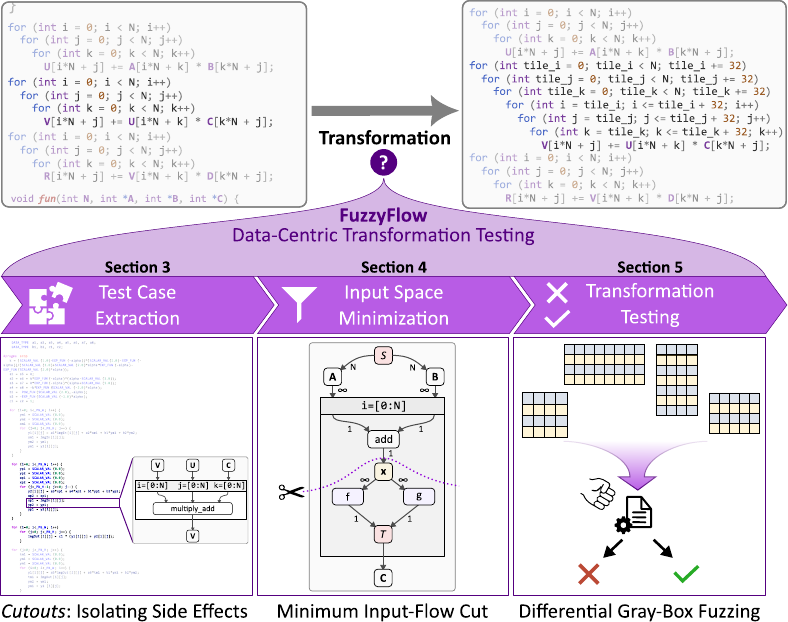}
  \vspace{-1.5em}
  \caption{Overview of FuzzyFlow.}
  \label{fig:overview}
  \vspace{-1em}
\end{figure}

\begin{figure*}
    \centering
    \includegraphics[width=\linewidth]{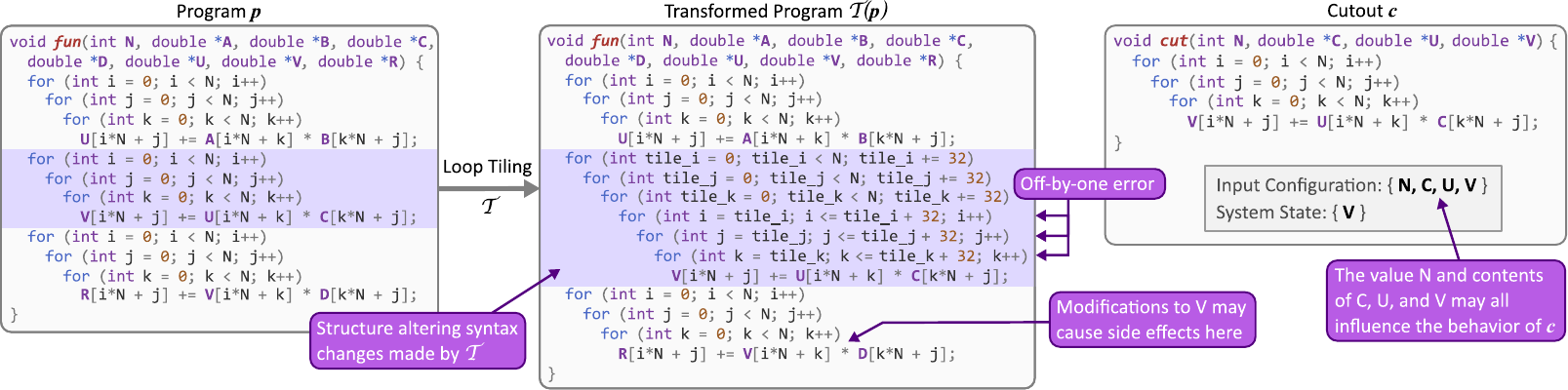}
    \vspace{-1.8em}
    \caption{Off-by-one error in tiling a matrix-matrix multiplication in a matrix chain multiplication $\boldsymbol{R = ((A \cdot B) \cdot C) \cdot D}$.}
    \label{fig:mm-tiling-demo}
\end{figure*}

Ideally a small test case that captures the changes made by a transformation can be extracted from the application, which allows equivalence checks to complete much faster.
However, capturing all semantic changes or \emph{side effects} such changes may introduce is often difficult due to program patterns such as pointer aliasing.


In this paper we specify the requirements for localized transformation validation and propose \emph{FuzzyFlow}:
a framework for testing transformations that leverages parametric dataflow program representations to extract minimal test cases capturing all transformation side effects, summarized in Fig.~\ref{fig:overview}.
By exploiting how dataflow languages expose the true data read-write set of each operation, it is possible to obtain the full system state of a test case for a given set of inputs and parameters.
Using differential testing~\cite{McKeeman1998DifferentialSoftware} we can then check that transformations do not alter this system state for the same set of input parameters.
To uncover input-dependent faulty behavior we employ gray-box fuzzing~\cite{Fioraldi2020AFLResearch} to sample different input configurations for testing.
We devise a dataflow-graph analysis technique that minimizes the test case input space at the cost of recomputation to reduce this testing effort.
With this approach we report any fault-inducing transformation instance and generate a fully reproducible, minimal test case including inputs that can aid in debugging transformations.

We demonstrate the effectiveness of this approach by providing a proof-of-concept implementation of FuzzyFlow that verifies custom transformations written in the DaCe optimization framework~\cite{Ben-Nun2019a}.
Using this implementation on a set of real-world scientific applications, we showcase how our technique enables fast optimization testing, and how the generated test cases enable simplified debugging on consumer workstations when optimizing HPC applications.

In summary, this paper makes the following contributions:

\begin{itemize}
    \item Requirements specification for localized transformation verification on general programs.
    \item Data-centric method for extracting \emph{minimal} test cases capturing all side effects of structural program changes to verify semantics preservation with differential testing.
    \item Graph-analytical approach to minimize test case inputs and reduce verification effort.
    \item A practical tool enabling up to 528 times faster optimization testing and debugging in real-world use cases.
\end{itemize}

\section{Localized Optimization Verification}\label{s:framework}
Optimizations are a special form of program \emph{transformations} that alter a program in an attempt to improve performance.
More generally, a program transformation $\mathcal{T}:\mathcal{P}\rightarrow\mathcal{P}$ is a function that takes a program $p \in \mathcal{P}$ and applies a syntactic change to arrive at a different program $p' = \mathcal{T}(p)$.
Syntactic changes made by a transformation may be contained to individual basic blocks, or they may alter the program structure.
In addition to that, a general transformation may operate \emph{within} program representations, such as with a source-to-source transformation, or it may transform a program from one representation into a different one.

Optimizations are a special subset of such transformations that typically operate within representations, often on a compiler's intermediate representation (IR), and aim to produce a program $p'$ that is semantically equivalent to the starting program $p$.
We say that two programs $p$ and $p'$ are semantically equivalent if, for a space of possible input data and parameters $\mathcal{S}_p$, it holds that $\forall s \in \mathcal{S}_p: p(s)=p'(s)$.
We denote this as $p \semeq p'$, and semantic inequality as $p \nsemeq p'$.

The tiling optimization shown in Fig.~\ref{fig:mm-tiling-demo} is an example of such a structure-altering source-to-source transformation.
The function shown computes a matrix chain multiplication $R = ((A \cdot B) \cdot C) \cdot D$ with four $N \times N$ matrices.
By tiling the loops that perform the second multiplication in the chain, the transformation aims to improve memory reuse but keep the product of the computation intact.
However, the transformation introduces a bug.
The loop conditions of the tiled loops are incorrectly using a \texttt{<=} comparison instead of \texttt{<}.
This off-by-one error changes the semantics of the program.
Executing the application would expose this problem, but if the multiplication is part of a larger application, that becomes costly.

Instead, the transformation can also be verified in the given example by only extracting the second matrix-matrix multiplication into a separate sub-program $c$, shown on the right in Fig.~\ref{fig:mm-tiling-demo}.
By checking that for any two $N \times N$ matrices $U$ and $C$, the results in $V$ are the same before and after tiling we can catch the erroneous transformation in less time.
This is possible because out of all the memory locations being written to inside of $c$, only the contents of $V$ can have an effect on the semantics of the third multiplication.
As such, we call $V$ the output or \textbf{system state} of $c$.
Similarly, the contents of the scalar value $N$, as well as the matrices $C$, $U$, and $V$ may influence the program behavior \emph{inside} of $c$ and consequently may influence the results of the computation.
We refer to this as the \textbf{input configuration} of $c$.

We call such a sub-program $c \subseteq p$ with a clear input configuration and system state a \textbf{cutout} of $p$.
With a given input configuration, a cutout $c$ can be treated as a stand-alone, executable program.
Because any change to $c$ that results in a change to the system state for the same input configuration indicates a change to the semantics of $c$, we can use cutouts to check for an optimizing transformation's preservation of semantics.

Specifically, for a given transformation $\mathcal{T}$ we want to find and extract a cutout $c$ from $p$ to capture all syntactic and structural changes made by $\mathcal{T}$ and then isolate possible side effects by capturing the input configuration and system state of $c$.
Using differential testing on $c$ and $\mathcal{T}(c)$ we can then check whether $c \semeq \mathcal{T}(c)$ holds by checking for changes to the system state.
Since the system state includes everything that can affect the remainder of the program, it then follows that $c \semeq \mathcal{T}(c) \implies p \semeq \mathcal{T}(p)$ and $c \nsemeq \mathcal{T}(c) \implies p \nsemeq \mathcal{T}(p)$.

Since large-scale algorithmic restructurings to $p$ may entail syntactic changes where the resulting cutout $c$ is not much smaller than the original application, this approach is particularly well suited for \emph{peephole} optimizations.
Peephole optimizations such as the tiling optimization demonstrated in Fig.~\ref{fig:mm-tiling-demo} apply smaller syntactic changes that can result in $c \lll p$, which leads to fast turnaround times during testing.
However, ensuring $c \lll p$ comes with a set of challenges.
Especially while still identifying any changes in program semantics and capturing all possible transformation side effects in the system state of $c$.
To address these challenges, the program representations used for extracting cutouts should fulfill a set of requirements.
We discuss these requirements and how different program representations fulfill them in the following subsections and provide an overview of them in Table~\ref{tab:requirements}.

\subsection{Generalization}\label{ss:generalization-req}
The first challenge with extracting and using cutouts to check for a transformation's correctness comes from the fact that some transformations may behave correctly for certain inputs, but not for others.
For example, even if the off-by-one error in the tiling optimization from Fig.~\ref{fig:mm-tiling-demo} is fixed, the transformation still contains a bug.
Specifically, due to the way the inner, tiled loops are bounded, the optimized program will cause out of bounds memory accesses for any inputs where $N$ is not a multiple of the tile size 32.

To detect this kind of input-dependent faulty behavior, a transformation's cutout should be evaluated under a series of different inputs and input sizes, or through symbolic execution.
However, not all possible inputs are considered valid inputs.
Certain values, such as pointers, may be used to indirectly address other memory, and changes to them can cause invalid memory accessing or undefined behavior.
Both in differential testing and symbolic execution, this can lead to a large number of false positives, making testing impractical.
As such, it is important that for a given cutout, we can analyze what input values may be modified to what extent to generalize test cases for many different inputs.

While difficult in SSA form representations such as LLVM-IR~\cite{Lattner2004}, MLIR~\cite{Lattner2021} dialects make it easy to detect when certain values are being used to reference other memory locations.
However, generalizing arbitrary cutouts to varying input \emph{sizes} is challenging in most program representations due to missing context.
For example, a pointer to a variable sized array tells us nothing about the size of the array without additional information.
This is demonstrated in the extracted cutout for our tiling optimization on the right side of Fig.~\ref{fig:mm-tiling-demo}.
By looking at the input configuration, the dependency between the scalar value $N$ and the size of the memory region pointed to by the pointer $C$ is opaque.
Both $N$ and the pointer are seemingly independent inputs, and evaluating it under different input sizes or symbolic execution would primarily cause segmentation faults or undefined behavior, causing false positives.

To avoid this, a representation used to extract cutouts should be \emph{parametric}.
This means that data containers are not expressed as pointers with unknown size, but instead with a known size and shape given by an expression $e$, where $e$ can be constant or dependent on certain program parameters.
In our tiling example that means the size of $C$ would be given with the expression $N \times N$, which re-establishes the relationship between the parameter $N$ and the data size, enabling testing with different sizes.

\begin{figure*}
    \centering
    \includegraphics[width=\linewidth]{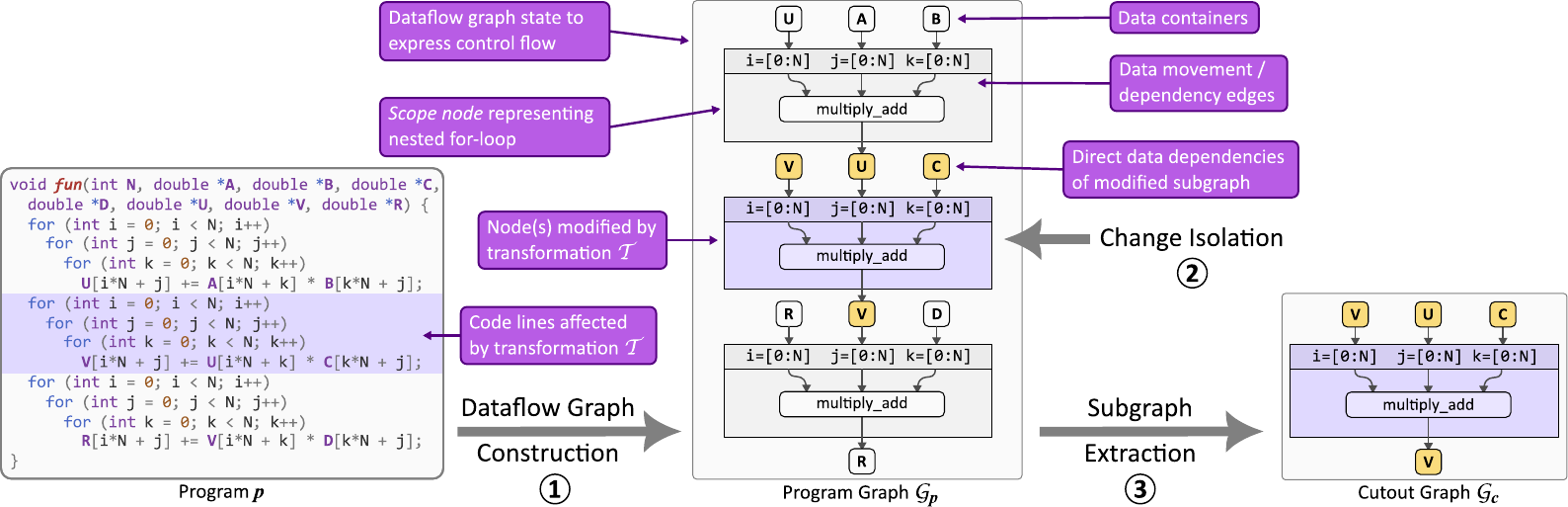}
    \vspace{-1.8em}
    \caption{Cutout extraction procedure for an arbitrary peephole optimization, demonstrated with a loop tiling transformation.}
    \label{fig:cutout-procedure}
\end{figure*}

\subsection{Side Effect Analysis}\label{ss:side-effect-analysis-req}
A further challenge arises when trying to determine the input configuration and system state for a program cutout based on the syntactic changes made by a transformation.
Since syntactic changes inside the cutout may affect semantics in the remainder of the application, it is important that the cutout's system state captures such changes.
In the tiling example from Fig.~\ref{fig:mm-tiling-demo}, the incorrect optimization causes different values to be computed for $V$ than before tiling, in turn affecting the results in $R$.
The change to the computation of $R$ is thus considered a \emph{side effect} of our transformation and the change to how $V$ is computed.
$V$ must therefore be part of the cutout's system state.

Correctly determining the cutout input configuration and system state consequently requires detecting any such side effects.
This form of side effect analysis is more challenging in some program representations than in others.
Even the change of a single variable value introduced through a syntax change may have side effects outside of the cutout and thus require said variable to be part of the system state.
The side effects arising from changes to a single register value or \emph{scalar} are easily exposed in static single assignment (SSA) form program representations, such as LLVM IR~\cite{Lattner2004}, as shown in Table~\ref{tab:requirements}.

\begin{table}
    \small
    \caption{Requirements for localized optimization testing.}
    \label{tab:requirements}
    \vspace{-1em}
    \setlength{\tabcolsep}{2.6pt}
    \begin{tabular}{@{}lccccc@{}}
    \toprule
    Requirements \footnotesize{\ding{220}} &
        \multicolumn{3}{c}{Side Effect Analysis} & \multicolumn{2}{c}{Generalization} \\
     \cmidrule(lr){2-4}\cmidrule(lr){5-6}
    Representation \rotatebox[origin=c]{-90}{\footnotesize{\ding{220}}} &
    Scalar &
    Memory &
    Sub-region &
    Inputs &
    Sizes \\
    \midrule
    Abstract Syntax Tree (AST) & \xmark & \xmark  & \xmark          & \xmark  & \xmark       \\
    SSA-Form~\cite{Lattner2004} & \cmark & \xmark  & \xmark          & \xmark  & \xmark       \\
    PDG~\cite{Ferrante1987TheOptimization}            & \cmark & \cmark  & \xmark          & \xmark  & \xmark       \\
    MLIR~\cite{Lattner2021}     & \cmark & \cmark  & \cmark\hdagger  & \cmark  & \xmark       \\
    Parametric Dataflow~\cite{Ben-Nun2019a, Kotsifakou2018HPVM}     & \cmark & \cmark  & \cmark          & \cmark  & \cmark       \\
    \bottomrule
    \end{tabular}
    {\begin{flushleft}\footnotesize $\dagger$ Constant sizes only.\end{flushleft}}
    \vspace{-1.5em}
\end{table}

However, it is harder to catch possible \emph{memory} side effects.
Program patterns such as pointer aliasing can mask the true effects of program statements.
Catching such aliasing or masking effects in source code, AST representations, or even SSA form representations is difficult without advanced and costly pointer analysis.
Representations like program dependence graphs (PDGs)~\cite{Ferrante1987TheOptimization} and other dataflow IRs expose the data dependencies between individual statements, simplifying this side effect analysis drastically.

Pointers also make it challenging to detect side effects related to accesses in specific memory \emph{sub-regions}.
This includes changes that affect overlapping memory regions or indexing changes which cause out of bounds accesses.
Given the difficulties of pointer analysis, these side effects are hard to detect in AST or SSA form program representations.
Even with the help of a PDG it is difficult to detect side effects related to overlapping memory regions.
Some built-in dialects in MLIR~\cite{Lattner2021} enable this form of sub-region or index analysis, as long as structs and arrays have constant sizes known during compilation.
Since parametric program representations keep the relationship between data containers and their size intact, they enable such analyses even for symbolic struct or array sizes where the concrete sizes are not statically known.

\subsection{Parametric Dataflow}\label{ss:bg-df}
The memory side effect analysis capabilities of dataflow IRs such as PDGs exist together with the generalization and sub-region analysis capabilities of parametric representations in \emph{parametric dataflow representations}.
This gives representations such as HPVM~\cite{Kotsifakou2018HPVM}, Naiad~\cite{Murray2013Naiad}, stateful dataflow multigraphs (SDFGs)~\cite{Ben-Nun2019a}, or Dryad~\cite{Isard2007Dryad} all the properties necessary to extract generalizeable program cutouts that isolate all side effects of program transformations.

We construct a reference implementation of our approach for the SDFG representation that is used by the optimization framework DaCe~\cite{Ben-Nun2019a}.
This framework has become a popular choice for optimizing scientific high performance computing applications from numerous fields, where engineers write purpose-built, custom optimizing transformations to get high performance gains out of their applications~\cite{Ivanov2021, Andersson2023ATransforms, Ziogas2022Deinsum:Algebra, Ben-Nun2022ProductivePython, Ziogas2019}.
Together with the ability to express arbitrary programs from Python, C, or Fortran, this makes the SDFG IR a good choice to demonstrate the effectiveness of our proposed approach.
However, the techniques outlined in this paper are generally applicable to any parametric program representation adhering to the requirements outlined in Table~\ref{tab:requirements}.

\paragraph{Stateful Dataflow Multigraphs}
Like most dataflow program representations, SDFGs are a graph-based program IR where graph nodes represent data containers and computations, and the edges connecting them represent data movement or dependencies.
Each data movement edge is annotated with the exact data subset being accessed in the connected computations.
Additionally, the graph is a \emph{hierarchical} dataflow graph, with different hierarchies nested inside one another.
Firstly, dataflow graphs are nested inside of states which are used to build a state machine that can express complex control flow.
Secondly, constructs like for-loops are expressed with special \emph{scope} nodes, where their loop body forms a nested dataflow graph inside of them.
An example of this IR can be seen in Fig.~\ref{fig:cutout-procedure}, which provides an overview of the cutout extraction procedure.

\section{Test Case Extraction}\label{s:cutouts}
Given a transformation $\mathcal{T}$ and program $p$, extracting a program cutout $c$ with the help of a graph-based IR that exposes dataflow can be thought of as extracting the dataflow subgraph which is modified by the transformation.
We split this procedure into three distinct steps summarized in Fig.~\ref{fig:cutout-procedure}:

\paragraph{\circled{1} Dataflow Graph Construction}
The first goal is to translate the program $p$ into its dataflow graph representation $\mathcal{G}_p$.
We can leverage DaCe~\cite{Ben-Nun2019a} for this purpose, which can construct parametric dataflow graphs from a variety of high level languages.

\paragraph{\circled{2} Change Isolation}
Next, the syntactic and structural changes made by a transformation need to be identified to get a starting point for what to extract into a cutout.
In a graph-based IR, this corresponds to identifying all graph nodes that have changed between the two program graphs $\mathcal{G}_p$ and $\mathcal{G}_{\mathcal{T}(p)}$.
If the change includes modified, added, or removed edges, both the edge source and destination nodes are considered to be modified.
We denote this set of modified nodes as $\Delta_{\mathcal{T}}$.
When treating transformations as black boxes, this change set has to be obtained through analyzing the difference between $\mathcal{G}_p$ and $\mathcal{G}_{\mathcal{T}(p)}$.

However, in some optimization frameworks, such as DaCe, transformations readily expose the set of changes made by reporting what specific program patterns they modify or by specifying rewrite rules.
Such white-box transformations thus do not require a difference analysis between $\mathcal{G}_p$ and $\mathcal{G}_{\mathcal{T}(p)}$.
Since the cutout $c$ where the transformation will be tested acts as an entirely separate program isolated from $p$, all effective changes made by a transformation are tested, even if the transformation incorrectly self-reports the change set.
If a transformation attempts to perform a change not captured by $c$, this will lead to a crash or exception since $c$ does not contain the corresponding element, thus exposing a problem.

\paragraph{\circled{3} Subgraph Extraction}
After identifying all affected dataflow graph nodes $\Delta_{\mathcal{T}}$, the corresponding subgraph can be extracted into a separate program.
We first construct a new, empty program graph $\mathcal{G}_c = \varnothing$ for our cutout $c$.
Now we can create a copy for each node $n \in \Delta_{\mathcal{T}}$ and add it to $\mathcal{G}_c$.
For every node $n$, we look at each incoming and outgoing edge connected to it in the dataflow graph.
If $n$ performs a computation, these connected edges in the dataflow graph expose what data containers are being accessed and at what indices or sub-regions.
If any of these data containers are not part of $\mathcal{G}_c$ yet, we copy them over as well.
This ensures that all direct data dependencies for the nodes affected by $\mathcal{T}$ are part of $\mathcal{G}_c$, including all properties of the corresponding data containers, such as their size.

When copying over data dependencies, we can additionally leverage the parametric properties of the dataflow representation to minimize the size of data containers required in $\mathcal{G}_c$.
Since parametric dataflow representations such as SDFGs allow us to analyze exactly what data indices or sub-regions are being accessed by each operation, we can avoid having to include entire data containers if only sub-regions are needed to capture dependencies.
For example, if a computation modified by $\mathcal{T}$ only accesses the indices 0 to 9 of an array \texttt{my\_arr} of size $N$, only the first 10 elements of \texttt{my\_arr} need to be included in $\mathcal{G}_c$.
This helps reduce both the size and memory consumption of our cutout $c$, consequently speeding up the testing process.

Fig.~\ref{fig:cutout-procedure} shows the cutout dataflow graph $\mathcal{G}_c$ that we can obtain by following this procedure based on the incorrect tiling transformation introduced in Section~\ref{s:framework} (see Fig.~\ref{fig:mm-tiling-demo}).

\begin{figure*}
    \centering
    \includegraphics[width=\linewidth]{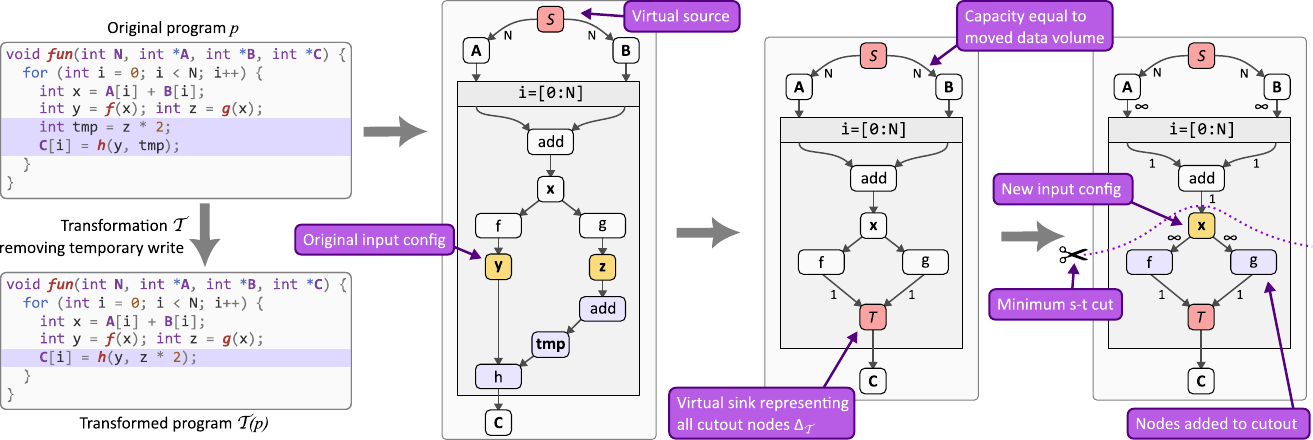}
    \vspace{-2.1em}
    \caption{Overview of the steps performed during the minimum input-flow cut process to reduce a cutout's input configuration.}
    \label{fig:state-minimization-example}
    \vspace{-.8em}
\end{figure*}

\subsection{Capturing Side Effects}\label{ss:capturing-side-effects}
Now that the dataflow graph $\mathcal{G}_c$ for our cutout $c$ contains everything directly affected by the $\mathcal{T}$, including all direct data dependencies, we need to make sure all possible side effects of $\mathcal{T}$ are captured.
This means we must correctly determine the \textbf{system state} the cutout $c$ leaves after its execution.
The system state consists of any data container or subset thereof written to inside the cutout, that may be read again elsewhere after the cutout's execution, if $c$ were placed back inside the original program $p$.
Any data container where this is the case can have an influence on the behavior of $p$ and consequently its results may affect semantics.

There are two ways in which data may be read again after $c$ was executed in the context of $p$, and that consequently need to be checked with two analyses:

\paragraph{External Data Analysis}
Any write operations $c$ performs to external or persistent data, such as program outputs or files on disk, always require the corresponding data container to be part of the system state.
In SDFGs, data containers helpfully have a property that indicates whether their allocation lifetime is \emph{transient} or not.
Any data container that is not marked as transient is not managed by the program and may persist, consequently leaving the chance to be read after the program has exited.

\paragraph{Program Flow Analysis}
Additionally, a write operation to a data container $d$ inside of $c$ requires $d$ to be part of the system state, if there is a read operation from $d$ inside of $\widebar{c} = p - c$ which is reachable from $c$.
To check for these side effects, we first analyze each write operation performed inside of $\mathcal{G}_c$ together with the sub-range or indices written.
We then perform a breadth first search (BFS) through the original program graph $\mathcal{G}_p$, starting at the nodes $n \in \Delta_\mathcal{T}$ that created the cutout.
For any read operation we encounter, we check if the data container being read is part of the set of write operations performed inside of $\mathcal{G}_c$, and if the read range or indices overlap with any of the written ranges or indices.
If yes, the data container is added to the system state of $c$.

\medskip
After these dataflow analyses, anything that is not marked as part of the system state does not have any influence on the semantics of $p$ and consequently does not need to be compared during the testing procedure.
There are some possible side effects that are not easily exposed even through dataflow representations, such as through the use of user-defined callbacks or libraries that may produce side effects.
Fortunately, their use can easily be detected, enabling clear warnings when transformations can produce unintended side effects that may not be caught.

\subsection{Determining Input Configurations}\label{ss:find-inputs}
To properly test a transformation's cutout under multiple circumstances, we must also extract everything that may be part of the \textbf{input configuration} for our cutout $c$.
The input configuration consists of all data containers that may already contain data before $c$ is executed, and may consequently influence the behavior of $c$.
We can obtain this input configuration with two analyses analogous to the ones performed for the system state:

In the external data analysis, we add the data containers of any read operations from external or persistent data inside of $c$ (data containers not marked as transient) to the input configuration.
In the subsequent program flow analysis, we analyze each read operation performed inside of $\mathcal{G}_c$ and take note of the read sub-ranges and indices.
By performing a reversed BFS starting at the nodes $n \in \Delta_\mathcal{T}$, we add any write operation to the input configuration if there is a corresponding read to an overlapping index inside of $\mathcal{G}_c$.
This uncovers any write operations that may reach the cutout through some execution path and consequently ensures that anything that may already contain data when the cutout is executed, is part of the input configuration.

\section{Minimizing Input Configurations}\label{s:minimizing-inputs}
For realistic cutouts, the size of the input configuration space can be very large.
Complex transformations in large applications can result in cutouts where the input configuration may consist of dozens of data containers.
During the differential testing to check whether a transformation is correct, we want to cover as much of this input configuration space as possible to make it unlikely that a fault inducing input goes undiscovered.
It is thus desireable to keep the size of this input configuration space as small as possible to reduce the testing effort required.

Fortunately, there are many situations where the input configuration can be reduced significantly from its initial size.
Consider for example the program shown on the left of Fig.~\ref{fig:state-minimization-example}.
A transformation attempts to subsume the computation \texttt{z * 2} into the function call to \texttt{h} to get rid of the write to the temporary data container \texttt{tmp}.
The cutout constructed through the procedure outlined before will include the multiplication and the function call to \texttt{h}.
The input configuration would consequently consist of both the data containers \texttt{y} and \texttt{z}.
However, by also including the function calls to \texttt{f} and \texttt{g} in the cutout, the input configuration only needs to contain the data container \texttt{x}.
This halves the input space and with it the effort required when testing different input configurations, at the cost of some additional computation.

\subsection{A Graph-Analytical Solution}\label{ss:a-graph-analytical-approach}
To automatically minimize the input configuration space, we devise a method to gradually expand the cutout with surrounding dataflow, until we find a cutout with a smaller input configuration space.
If the input space cannot be further minimized, the original cutout is used.
Since the testing effort only depends on the size of the input configuration space, we can refrain from minimizing the system state of the cutout.
Consequently, we can ignore anything that happens in the program after the cutout, and only need to consider including more of the cutout's predecessor nodes in the graph.

Since the edges in a dataflow graph represent data movement, they have a certain data \emph{volume} associated with them, which is being moved or accessed across them.
For all data containers that are part of our input configuration, that data movement volume used to access them thus corresponds to the amount of data needed to sample a single input configuration.
We can think of this data movement volume as the \emph{capacity} of each data movement edge.

This can be exploited to reformulate finding the optimal solution for minimizing the input configuration space as finding the minimum s-t cut in the dataflow graph between the start of the program and our cutout.
In a flow graph, the minimum s-t cut refers to a partitioning of the graph between a source node and a sink node into two disjoint components, such that the sum of edge capacities crossing between the two components is minimal.
In our context, a cut satisfying that property translates to a cut through the dataflow graph that minimizes the volume of data being moved between components.
This minimal flow of input data into the cutout component thus minimizes the size of the input configuration space.

\subsection{Minimum Input-Flow Cut}\label{ss:minimum-input-flow}
The procedure for finding a cut between the start of the program in $\mathcal{G}_p$ and our cutout graph $\mathcal{G}_c$ that minimizes this \emph{input-flow} can be split into two separate phases:

\paragraph{Preparation}
Since the dataflow graph $\mathcal{G}_p$ may have more than one source node, i.e., nodes without incoming edges, there is no unique start node that can be used for the minimum cut.
To address this, we insert a dummy source node $S$ into $\mathcal{G}_p$ and connect it to each node which does not have any incoming edges.
If the node in question is a data node, we give the edge a capacity equal to the size of the data container that node represents.
Similarly, the cutout may consist of more than one graph node, meaning that there is no unique sink in $\mathcal{G}_p$ to represent the cutout either.
Instead, we insert a dummy sink node $T$ into $\mathcal{G}_p$ to represent $\mathcal{G}_c$ as one node.

Next, we create edges between $S$ and each data node that accesses external or persistent data.
The capacity of those edges is set to the size of the data containers being accessed by each node.
All other incoming edges to those data nodes have their capacity set to $\infty$.
This is necessary since each access to external or persistent data is always part of the input configuration, as discussed in Sec.~\ref{ss:find-inputs}.

We can now look at each data node inside of $\mathcal{G}_c$ that belongs to the initial input configuration of the cutout.
For each such node, we redirect all incoming edges such that their destination is set to the sink node $T$ instead.
Additionally, the capacity of these edges is set to the volume moved across them, which we can derive by looking at the data sub-range or indices being accessed through them.

To complete the process of connecting our virtual sink node $T$, we now look at the outgoing edges for each node in $\mathcal{G}_c$.
For any such edge where the destination lies outside of $\mathcal{G}_c$, we have two situations to consider: either the destination has a path back to a node which lies inside of $\mathcal{G}_c$, or it does not.
If there is no path back to $\mathcal{G}_c$, then we simply change its source to $T$.
However, if there is a bath back to $\mathcal{G}_c$, we can redirect the edge to lead from $S$ to $T$ with a capacity of 0.
This is possible because any data the edge accesses will already be part of the cutout, and consequently does not need to be added to the input configuration.
The data movement volume across it is consequently considered 'free' in the context of our minimum input-flow cut.
After making these connections, we remove all nodes in $\mathcal{G}_c$ and any connected edges from $\mathcal{G}_p$.

Finally, we run over all data nodes in $\mathcal{G}_p$ and set the capacity of all of their outgoing edges to $\infty$.
We do this since a cut over one of those edges would sever a data dependency without adding the corresponding data node to the cutout component.
Instead, we want a potential cut to happen before the data node.
This preparation process is demonstrated on the transformation removing temporary writes shown in Fig.~\ref{fig:state-minimization-example}.

\paragraph{Minimum S-T Cut}
In the second phase, we use the program graph $\mathcal{G}_p$ prepared in the previous phase to solve the minimum s-t cut problem and find a cutout with a reduced input configuration space.
The max-flow min-cut theorem~\cite{Dantzig1955OnNetworks} states that the maximum flow through a flow graph (or network) is given by the minimum s-t cut through the graph.
This means that we can use an existing efficient implementation of the Ford-Fulkerson algorithm~\cite{Ford1956MaximalNetwork} for finding the maximum flow from $S$ to $T$, such as Edmonds-Karp~\cite{Edmonds1972TheoreticalProblems}, to determine our minimum cut.

Up until now, the capacity of each edge may still be given as a symbolic expression due to the parametric nature of the dataflow representation.
Since such maximum flow algorithms don't work with symbolic capacities, we concretize the symbol values at this point in the process with constant values that may be provided by the user.

For $\mathcal{G}_p$ with the set of nodes $V$ and the set of edges $E$, the Edmonds-Karp maximum flow algorithm finds the minimum cut with a time complexity of $\mathcal{O}(|E|^2|V|)$.
Given the minimum s-t cut, we can now extend the cutout by everything that resides in the same component of $T$ and can reach $T$, to arrive at a cutout with the smallest possible input configuration space.

\section{Differential Testing}\label{s:fuzzing}
With an obtained minimal cutout $c$ that captures the changes and side effects of a transformation $\mathcal{T}$, testing whether $\mathcal{T}$ preserves semantics can now be done by checking that $c \semeq \mathcal{T}(c)$ holds.
If we recall from Sec.~\ref{s:framework}, for a cutout $c$ with a space of possible input data and parameters $\mathcal{S}_c$, $c \semeq \mathcal{T}(c)$ can be shown by showing:
\[
\forall s \in \mathcal{S}_c: c(s) = c'(s),~\mathrm{where}~c' = \mathcal{T}(c)
\]

However, even after minimizing the size of the input configuration space, the number of possible input configurations $|\mathcal{S}_c|$ may still be high.
Even for a trivially small cutout that adds two 32-bit integers together, $|\mathcal{S}|$ is equal to $2^{64}$.
Verifying each of these input configurations would take an impractically long time or may be entirely infeasible.

\paragraph{Symbolic Execution}\label{ss:symbolic-execution}
A popular approach to avoid having to run all possible input configurations is to employ symbolic execution or concolic testing~\cite{Hathhorn2015DefiningC, Cadar2008}.
Symbolic execution is a widely used technique that has found a strong foothold in software and security testing.
However, symbolic execution faces a few challenges that make it impractical for the context of scientific high performance computing applications.

Specifically, symbolic execution struggles with anything where it does not have direct source knowledge, such as library or system calls, and may suffer from state space explosion in the context of nested loops~\cite{Baldoni2019ATechniques}.
Unfortunately, both of these challenging situations are encountered regularly in high performance, scientific code.
Substitution of computations with architecture-specific intrinsics or nested loop optimizations such as tiling are often the subject of program optimizations in supercomputing contexts.

In addition to that, most scientific applications heavily rely on floating point arithmetic in their computations.
While there are techniques to perform symbolic execution in the context of floating point operations, they currently still lack the robustness to efficiently and reliably catch issues.
Most widely used frameworks such as KLEE~\cite{Cadar2008} thus do not support floating point symbols yet.
These limitations make symbolic execution a less desirable technique in the context of scientific HPC applications.

\subsection{Differential Fuzzing}\label{ss:transformation-verification}
Instead of using symbolic execution, we can use fuzzing~\cite{Manes2018} to randomly sample input configurations from $\mathcal{S}_c$ and evaluating them over a series of $t \lll |\mathcal{S}_c|$ trials.
This still avoids having to test each individual input configuration, and instead samples inputs in such a way that the likelihood of finding issues is maximized.

In each trial, an input configuration $s \in \mathcal{S}_c$ is run through both the original cutout $c$ and $\mathcal{T}(c)$.
For each such trial, the system state of the two cutouts after execution is compared and any system state variation $c(s) \neq c'(s)$ labels the transformation as invalid.
Specifically, a change in the system state is reported if either the transformed program $c'$ crashes or hangs while $c$ does not, or the numerical results produced differ by more than a given threshold $t_\Delta$.
The threshold $t_\Delta$ can be configured by the user\footnote{For this paper we use $1e^{-5}$.}, and if $t_\Delta$ is not provided or set to $0$, we instead test for bit-wise equality.

\paragraph{Gray-Box Fuzzing}\label{ss:sampling-inputs}
To achieve a high probability of detecting a possible change in semantics, it is imperative that the configurations sampled over $t$ trials cover the input configuration space well.
This can be achieved by sampling uniformly at random from the space of all possible values for each input parameter, and doing so for all possible data types where they are not explicitly stated.
This form of uniform sampling fully covers the input space, but comes with two additional challenges.
Firstly, uniform sampling may lead to many uninteresting crashes, where both $c$ and $\mathcal{T}(c)$ do not execute successfully.
Secondly, the program may take different program paths based on specific input values, in which case evaluating all possible program paths becomes statistically unlikely through uniform sampling.

To address the first challenge, we instead perform a form of gray-box fuzzing, meaning that we perform some amount of static analysis on both $p$ and $c$ to derive constraints for sampling.
By deriving constraints for certain input parameters based on the context in which they are used we reduce the number of uninteresting crashes while simultaneously further reducing the space of possible input configurations.
This in turn increases the likelihood with which a possible change in semantics can be uncovered.

We can perform two kinds of constraint analyses.
First, we check if any input parameters are used to access other data containers inside of our cutout's dataflow graph $\mathcal{G}_c$.
This is again where the parametric nature of the representation can help us by letting us analyze the sub-regions and indices accessed by each data movement edge.
Any parameter that is used to access inside a data container is bound to the interval $[0, Dim_{max}]$, where $Dim_{max}$ is the size of the data container in that specific dimension.
For example, for an $N \times M$ matrix $A$, the access $A[i, j]$ would constrain $i$ to the interval $[0, N]$ and $j$ to $[0, M]$.
A second analysis is performed on the original program's dataflow graph $\mathcal{G}_p$ to determine if any program context constrains parameters to a specific range.
Of particular interest here are loop iteration variables that may be constrained to certain loop bounds, if the cutout was taken from inside one or more loops.

In addition to these derived constraints, an engineer may further constrain the testing process by providing custom constraints.
This can be useful if it is known through domain knowledge that certain program parameters are always observed to have certain ranges.
Finally, since a data container can never have a size of $\leq 0$, any parameter that is used to determine the size of a data container is only sampled in the range $[1, Size_{max}]$, where $Size_{max}$ can be configured arbitrarily.

\paragraph{Coverage-Guided Fuzzing}\label{ss:coverage-guided}
To address the second challenge with uniform sampling, we can additionally use the obtained constraints to perform coverage-guided fuzzing.
Here, the code is instrumented and program path coverage is recorded for each fuzzing trial.
The fuzzer then attempts to mutate the sampled inputs in such a way that new program paths are taken in an attempt to maximize program coverage, consequently increasing the likelihood of finding bugs.

To enable this, an SDFG-based cutout $c$ and its transformed counterpart $\mathcal{T}(c)$ can be turned back into C++ programs together with a small auto-generated harness that calls both programs and compares their outputs.
If the cutout outputs differ, the harness causes a segmentation fault.
This allows the use of existing coverage-guided fuzzers such as AFL++~\cite{Fioraldi2020AFLResearch} or other advanced fuzzing tools.
While the powerful constraint analyses discussed before may not be performed this way, this allows us to profit both from the localized testing capabilities gained through cutouts, as well as the years of research into advanced fuzzing techniques~\cite{Manes2018}.

\section{Case Studies}\label{s:evaluation}
We demonstrate the use of FuzzyFlow in a real-world setting with case studies from working with and optimizing scientific HPC applications.
To illustrate how the proposed technique helps test and debug transformations in HPC contexts on consumer workstations, we run each test on a consumer-grade system with a 10-core Intel i9-7900X CPU at 4.5 GHz and 32 GB of RAM.
We use Python 3.9, a development build of DaCe 0.14.2, GCC 10.3 and AFL++ 4.06a.

\subsection{Minimizing Input Configurations}
Machine learning workloads are an increasingly important part of the HPC landscape, consuming large amounts of compute resources for applications spanning diverse fields.
With this case study we demonstrate how our minimum input-flow cut algorithm can significantly reduce the input space for extracted test cases while optimizing a machine learning application.
For this purpose we select the encoder layer from the natural language model BERT~\cite{Devlin2019}.
This type of Transformer~\cite{Vaswani2017a} is a widely used neural network, where even pre-trained models take hours to tune in large-scale distributed environments. 

The Multi-Head Attention (MHA) in BERT's encoder layer contains a series of nested loops which perform a variety of element-wise operations, tensor contractions, and normalizations.
Using a built-in transformation in DaCe, we can optimize most of these nested loops by vectorizing them.
The transformation tiles the loops by the chosen vector size (4 by default) and augments the computations with vector instructions.
Using FuzzyFlow, we test each instance of the transformation before applying to ensure program semantics remain unchanged.
We optimize the application for use with the $\textsc{BERT}_\textsc{LARGE}$~\cite{Devlin2019} model size (batch size $B=8$, attention heads $H=16$, hidden size $N=1024$, input/output sequence length $SM=512$, embedding size $emb=4096$, and projection size $P=\frac{I}{H}=64$), which takes 12.1 seconds to run when accelerating BLAS operations with Intel MKL~\cite{mkl}.

\begin{figure}
    \centering
    \includegraphics[width=.9\linewidth]{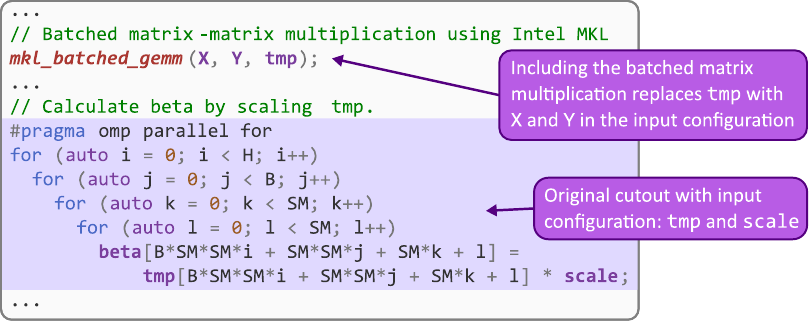}
    \vspace{-0.4em}
    \caption{Input space reduction for a loop nest cutout.}
    \vspace{-1em}
    \label{fig:input-config-bert}
\end{figure}

One instance of such a loop nest which performs the scaling of a tensor \texttt{tmp} is shown in Fig.~\ref{fig:input-config-bert}.
The resulting cutout capturing all syntactic changes from vectorizing the loop nest is highlighted in the figure and contains only the loop nest itself.
The input configuration of this cutout initially consists of the $B \times H \times SM \times SM$ tensor \texttt{tmp} and the scalar value \texttt{scale}.

Using the minimum input-flow cut algorithm to reduce the size of the input space, we arrive at a larger cutout, which also includes the batched matrix-matrix multiplication that computes \texttt{tmp}.
By including the computation of \texttt{tmp} it is no longer required as an input, and is instead replaced by the two $P \times B \times H \times SM$ tensors \texttt{A} and \texttt{B}.
With the chosen program parameters this \textbf{reduces the input configuration by 75\%}, consequently increasing test coverage.

The reduction in the input space additionally causes a 2$\times$ speedup in sampling input values and checking system state equivalence.
Using AFL++ to perform coverage-guided fuzzing, this allows running an average of 43.7 fuzzing trials per second, which is \textbf{528 times faster} compared to testing the transformation by running the entire application.
When testing for different model sizes, AFL++ takes an average of 157 trials per vectorization optimization to discover that the transformation's correctness depends on specific input data sizes.
By using our own gray-box fuzzing with parameter constraint analyses it only takes an average of 1 trial to uncover this input-dependent correctness, but at the cost of longer trial runs.
The longer trial runs are primarily due to architectural setup and teardown penalties such as copying data between processes, but this overhead can be optimized by adopting strategies found in existing state of the art fuzzers such as AFL++.

\subsection{From Multi-Node to Single-Node}
Another category of increasingly important machine learning workloads are graph neural networks (GNNs)~\cite{Besta2022ParallelAnalysis, Chami2022MachineTaxonomy, Wu2021ANetworks} such as Vanilla Attention~\cite{Velickovic2017GraphNetworks}, which lends itself well for running in a distributed setting.
Vanilla Attention performs a Sampled Dense-Dense Matrix Multiplication (SDDMM) in each layer during forward propagation, which is the target of many optimization efforts due to its poor data locality~\cite{Yu2021ExploitingGPUs, Nisa2018SampledLearning}.
However, testing optimizations performed on SDDMM for correctness or debugging them can be challenging, especially in a distributed scenario.
Tests that span multiple compute nodes are expensive to run, and not all applications can trivially be reduced to single-node contexts since behavior may vary between ranks or different communicator sizes.

\begin{figure}
    \centering
    \includegraphics[width=.9\linewidth]{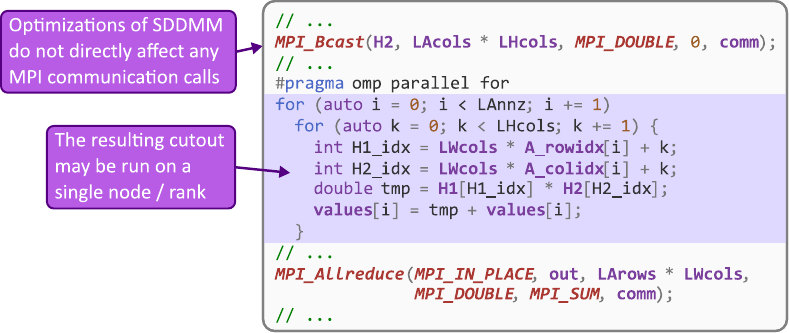}
    \vspace{-1em}
    \caption{Cutout extraction enables testing on a single node.}
    \label{fig:mpi-example}
    \vspace{-1.5em}
\end{figure}

By using the cutout-based testing process in FuzzyFlow we can effectively reduce test cases to a single rank for any optimizations that do not directly affect communication operations.
An example of this is shown in Fig.~\ref{fig:mpi-example}, where extracting a cutout for SDDMM in Vanilla Attention allows testing on a single node.
Since cutouts only include the direct data dependencies of the optimizing changes and detect side effects through changes in the cutout's system state, communication does not need to be included in a cutout if it is not directly modified.
Any data received through collectives is subsequently exposed as regular data parameters to the cutout and subjected to differential fuzzing.

\subsection{NPBench}
Due to the increasing use of DaCe for program optimizations we test each of its built-in optimizations\footnote{Some transformation requiring special hardware, such as FPGAs, are omitted.} on a set of micro-benchmarks.
We use the benchmark suite NPBench~\cite{Ziogas2021}, which contains a total of 52 benchmark programs from various application domains.
For each benchmark application, we test each individual instance where any of the built-in optimizations can be applied.
Most of the resulting \emph{3,280 transformation instances} pass this testing procedure.
However, we identify a total of \textbf{6 transformations} containing bugs and one transformation where correctness depends on specific combinations of input parameters and options passed to the transformations.
We summarize our findings in Table~\ref{tab:eval-overview}.

\subsection{Optimizing Weather Forecasts}
The cloud microphysics scheme (CLOUDSC) is part of the European Centre for Medium-Range Weather Forecasts’ (ECMWF) Integrated Forecasting System (IFS)~\cite{cloudsc}, and its Fortran implementation spans 3,163 lines of code.
We worked with a group of engineers tasked with optimizing this application using DaCe by primarily exploiting more parallelism and porting the application to accelerators.
In this process, the engineers wrote a series of transformations which were then applied to the application automatically.

While these transformations improved performance significantly, they also caused numerical errors in the computations.
Due to the size of the application and the large number of transformations applied, identifying which transformation instances caused errors and then building minimal test cases to debug them was a time-consuming process.
Just identifying a \emph{single} transformation bug that incorrectly extracted a GPU kernel took one engineer \emph{over 16 hours}.
This was only possible by reducing the problem size, which often is not an option.

With this case study we demonstrate how the use of FuzzyFlow could help in such situations.
Using FuzzyFlow we test each instance where the transformations used can be applied to the application over 100 trials.
We do this to isolate which transformations contain bugs and find fault-exposing test cases to assist in debugging.
We identify 3 transformations that cause semantic changes when applied to CLOUDSC at various stages of the optimization process.

\paragraph{Extract GPU Kernels}
We first test a custom transformation written by the engineers while optimizing CLOUDSC, which automatically extracts GPU kernels from the application.
It does this by identifying routines that may be run as GPU kernels and then generating CUDA code for those routines and inserting the necessary boilerplate such as kernel launch calls and data copies between the host and the GPU.
We test a total of 62 instances of this transformation on CLOUDSC, out of which we identify \textbf{48 instances} that alter program semantics.

\begin{figure}
    \centering
    \includegraphics[width=\linewidth]{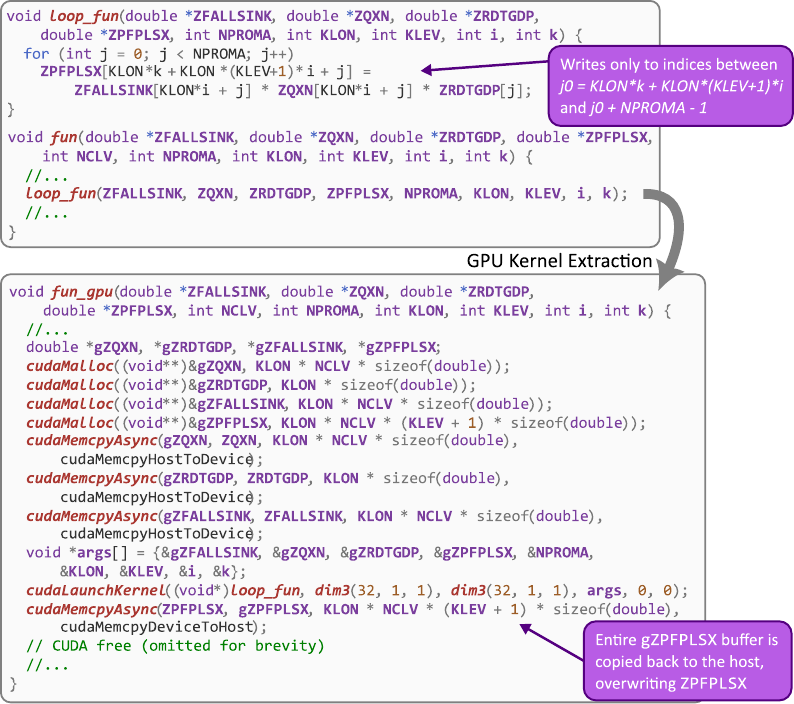}
    \vspace{-2em}
    \caption{GPU kernel extraction overwriting host data.}
    \vspace{-1.4em}
    \label{fig:gpu-transform-bug}
\end{figure}

One of the failing test cases generated is shown in Fig.~\ref{fig:gpu-transform-bug}.
A closer look reveals that the problem lies in the fact that the transformation generates data copies for the entire data containers touched by extracted GPU kernels, even if the kernel only reads or writes a subset of the data.
If the data written to by the kernel is not also first copied on to the GPU in its entirety, this causes garbage values to be copied back to the host, potentially overwriting existing computation results.
This test case took only one trial and 43 seconds to identify the transformation instance as invalid, and all other 48 invalid instances were similarly uncovered after 1-2 fuzzing trials each.
This indicates that at least 16 person-hours could have been saved in identifying this issue through the use of FuzzyFlow.

\begin{table}
    \small
    \begin{center}
    \caption{DaCe transformation bugs uncovered by FuzzyFlow.}
    \vspace{-1em}
    \label{tab:eval-overview}
    \setlength{\tabcolsep}{2.6pt}
    \begin{tabular}{@{}lc@{}}
        \toprule
        Transformation               & Failure \\
        \midrule
        \textbf{\texttt{BufferTiling}:} Tiles buffers between loops & {\normalsize\xmark} \\
        \textbf{\texttt{TaskletFusion}:} Removes temporary writes & {\normalsize\xmark} \\
        \textbf{\texttt{Vectorization}:} Vectorizes loops & \raisebox{.3ex}{\danger} \\
        \textbf{\texttt{MapExpansion}:} Removes collapsing from parallel nested loops & {\footnotesize\faHammer} \\
        \textbf{\texttt{MapReduceFusion}:} Removes intermediate buffers for reductions & {\footnotesize\faHammer} \\
        \textbf{\texttt{StateAssignElimination}:} Program simplification & {\footnotesize\faHammer} \\
        \textbf{\texttt{SymbolAliasPromotion}:} Program simplification & {\footnotesize\faHammer} \\
        \bottomrule
    \end{tabular}
    \end{center}
    {\footnotesize {\normalsize\xmark}~Change in semantics\quad \raisebox{.3ex}{\danger} Input dependent\quad {\footnotesize\faHammer}~Generates invalid code}
    \vspace{-.6em}
\end{table}

\paragraph{Loop Unrolling}
A second custom transformation where we discover bugs using FuzzyFlow performs loop unrolling.
We test a total of 19 instances of this transformation on CLOUDSC, uncovering \textbf{one instance} where semantics are altered through applying the transformation.
By examining the extracted test case it is discovered that the loop being unrolled has a negative loop step, iterating from \texttt{i=4} down to \texttt{i=1} with a step of \texttt{-1}, where \texttt{i} is the iteration variable.
While this means that the loop is executed 4 times, the transformation incorrectly unrolls the loop by only creating 2 loop body instances.

\paragraph{Write Elimination}
The last faulty transformation identified is a built-in optimization in DaCe which eliminates temporary write operations between computations.
We run 136 instances of this transformation on CLOUDSC through FuzzyFlow, which uncovers \textbf{one instance} where the transformation causes a change in semantics.
By subsuming one computation into the other in this specific instance, the test case reveals that the transformation removes an intermediate write to a data container which was marked as part of the test cutout's system state.
This means that the intermediate value is read again at a later point in the application, thus implying that the write to that data container should not be removed.

\section{Discussion}\label{s:discussion}
While the proof-of-concept implementation used to demonstrate the effectiveness of the workflow enabled by FuzzyFlow is built on top of DaCe, the core elements of the approach are not reliant on any specific optimization framework.
The core techniques introduced that facilitate this localized transformation testing approach consist of three dataflow graph analyses:
\begin{itemize}
    \item Identifying all side-effects of arbitrary, structure-altering program changes.
    \item Reducing the input space of extracted sub-programs through a min-cut formulation on the dataflow graph.
    \item Deriving input constraints on extracted test cases to reduce false positives during differential fuzzing.
\end{itemize}
These analyses are facilitated by dataflow languages through exposing true data dependencies on multiple levels (see Sec.~\ref{ss:side-effect-analysis-req}).
As such, the outlined techniques can readily be applied to other dataflow representations, such as HPVM~\cite{Kotsifakou2018HPVM}, Naiad~\cite{Murray2013Naiad}, or Dryad~\cite{Isard2007Dryad}.
Optimization frameworks utilizing different program representations may follow the outlined workflow for localized optimization testing if their internal representation can be mapped to such a parametric dataflow representation.

\subsection{Limitations}\label{ss:limitations}
There are two classes of transformation bugs that may go undetected despite the use of a dataflow representation.
Firstly, if a bug is only observed with a very small number of possible input values, probabilistic fuzzing may not detect the issue.
As discussed in Sec.~\ref{ss:coverage-guided}, coverage-guided fuzzing attempts to mitigate that risk by trying sample inputs that explore different program paths.

Secondly, if a transformation manipulates user-defined callbacks or library calls that may carry side effects, a dataflow representation does not allow us to capture all possible side effects.
However, it is possible to detect their presence inside of a transformation's change set and provide adequate warnings in those situations (see Sec.~\ref{ss:capturing-side-effects}).

\section{Related Work}\label{s:relwork}
Almost half a century's worth of research has been dedicated to proving the validity of compilers~\cite{Chen2021ATesting}.
Several techniques have been developed with the goal of verifying the transformations made by an optimizing compiler, and there are different approaches for reducing programs to smaller test cases for specific situations.

\paragraph{Compiler Optimization Verification}
Works on verifying the preservation of semantics through compilers and their optimizing transformations have taken several different approaches.
Some, such as CompCert~\cite{Leroy2006FormalAssistant, Leroy2009FormalCompiler} or CakeML~\cite{Kumar2014CakeML:ML}, take the approach of verifying the compiler as a whole, guaranteeing that any transformations and translations performed preserve program semantics.
Others observe the compilation or optimization process and check the program itself for equivalence before and after each separate optimization pass in a process called translation verification~\cite{Pnueli1998, Pnueli1998TheCVT, Pnueli2006TranslationOptimizations}.
PolyCheck~\cite{Bao2016PolyCheck:Programs} augments programs with light-weight checker codes to automatically verify loop-based transformations of affine programs.
While these checker code acts similar to program cutouts in avoiding whole-program testing, data-centric program cutouts allow testing for arbitrary programs outside the affine space.
VOC~\cite{Zuck2002VOC:Compilers, Zuck2005TranslationTransformations} and a translation verification infrastructure proposed by Necula~\cite{Necula2000} use simulation relations between the IR of a program before and after optimization steps to infer whether the two are semantically equivalent.
Rival~\cite{Rival2004SymbolicCompilation} follows a similar approach, representing changes in the IR with symbolic transfer functions, and in CoVaC~\cite{Zaks2008}, equivalence checks are similarly performed with a comparison system between the transformed and original version of a program.
Morpheus~\cite{Mansky2016SpecifyingGraphs} and the verification framework VeriF-OPT~\cite{Mansky2014} use a formal specification language to express transformations as rewrites on control flow graphs with side conditions to verify individual transformations.
Farzan and Nicolet~\cite{Farzan2019ModularLoops} formally verify the correctness of parallelization for read-only loop-nest by looking for multi-dimensional homomorphisms between loop-nests before and after parallelization.
UC-KLEE~\cite{Ramos2011PracticalCode} synthesizes fixed-size test inputs for C functions before and after optimization and compares their behavior with respect to exit codes and detected escaping memory locations.
However, like most techniques, this is often impractical for large, scientific applications.

With our approach we remove the need for modifying the compiler or the associated translation rules by trading in formal correctness proofs for differential fuzzing.
Additionally, FuzzyFlow enables the testing of arbitrary, structure altering transformations, while most formal verification approaches discussed are limited to certain program classes (e.g., affine, SCoP), or intraprocedural, structure preserving, or single-loop optimizations.

\paragraph{Test Case Extraction}
Many test case extraction approaches and frameworks are based on a technique called delta-debugging~\cite{Zeller2002SimplifyingInput, Zeller1999}, where specific failure inducing changes are isolated through a binary search on the total set of changes.
This is a technique employed and further developed by many works, including LLVM's Bugpoint~\cite{bugpoint}, Bugfind~\cite{Caron1990Bugfind:Compilers}, vpoiso~\cite{Whalley1994AutomaticErrors}, Perses~\cite{Sun2018Perses:Reduction}, HDD~\cite{Misherghi2006HDD:Debugging}, and C-Reduce~\cite{Regehr2012}.

However, these techniques typically answer the question of \emph{where} a bug resides and aim to provide a minimal test case for.
When it is not known if an error is present, a failure inducing input is usually not available, which makes these techniques suboptimal for asking \emph{if} an error is present.

Program slicing~\cite{Xu2005ASlicing, Weiser1984ProgramSlicing} techniques instead aim to decompose programs to obtain all statements related to a specific computation -- the slicing criterium.
Many approaches that slice programs statically~\cite{Ernst1994PracticalCode, Ottenstein1984TheEnvironment, Horwitz1990InterproceduralGraphs} perform the decomposition by analyzing a program dependence graph (PDG)~\cite{Ferrante1987TheOptimization} or iterations thereof, such as value dependence graphs (VDGs)~\cite{Weise1994ValueGraphs}.

By relying on representations such as PDGs, most static slicing approaches need to perform overly conservative side effect analyses that may lead to large reductions.
Through the use of parametric dataflow representations we are able to construct much smaller, side effect free program cutouts, with the added benefit of test cases remaining generalizeable to different inputs and input sizes.

\paragraph{Fuzzing}
The body of research around fuzzing, particularly in information security, is enormous~\cite{Manes2018}, with many well-known and established fuzzing frameworks like AFL~\cite{americanfuzzylop} and AFL++~\cite{Fioraldi2020AFLResearch}, funfuzz~\cite{funfuzz}, or BFF~\cite{certbff}.
The technique is also applied to testing and verifying compiler correctness~\cite{Boujarwah1997CompilerAssessment}, for example in Csmith~\cite{Yang2011FindingCompilers}, which employs black-box fuzzing to generate random test cases for C compilers to trigger unwanted behavior.
Linding~\cite{Lindig2005RandomConventions} and Sheridan~\cite{Sheridan2007} use a similar technique to generate random C programs for exposing inconsistencies in C compilers.

By extracting small, side effect free program cutouts to test the implications of a transformation, this allows us to build on the years of experience in fuzzing and the contributions made by each of these approaches.

\section{Conclusion}\label{s:conclusion}
We develop an approach that leverages parametric dataflow representations to extract side effect free program \emph{cutouts} to capture the changes made by program transformations and test them for correctness.
By solving a minimum s-t cut graph problem on the dataflow graph, we minimize the number of possible input configurations for extracted cutouts to speed up testing and improve test coverage.
Using gray-box differential fuzzing, program transformations are tested under various different program conditions to uncover even input dependent faulty behavior.
With a reference implementation built on the optimization framework DaCe, we demonstrate how this technique enables up to 528 times faster program optimization testing in real-world applications when compared to traditional techniques.
The fully reproducible, minimal test cases with fault-inducing inputs generated when a transformation bug is found additionally allow debugging of optimizations for supercomputer-scale applications on consumer workstations.

This novel transformation test case extraction approach opens the door for further advancements in both automatic program optimization testing and debugging.
By applying fuzzing to the structure of program cutouts a transformation is applied to, or the parameters for a given transformation, such as the tile size in a tiling optimization, transformations can be tested under even more varying conditions.
Additionally, the properties of dataflow representations in program cutouts may be exploited to further assist in transformation debugging, by highlighting exactly at which point along the dataflow path values begin to diverge after a faulty transformation was applied.
We have demonstrated how the techniques discussed in this paper enable more interactive optimization testing and debugging workflows, allowing performance engineers to save costly person-hours and distributed computing resources.

\begin{acks}
This project received funding from the European Research Council (ERC) grant PSAP, grant agreement No. 101002047, and the European Union's Horizon Europe programme DEEP-SEA, grant agreement No. 955606.
P.S. and T.B.N. are supported by the Swiss National Science Foundation (Ambizione Project No. 185778).
\end{acks}

\bibliographystyle{ACM-Reference-Format}

\end{document}